# Designing of RF Single Balanced Mixer with a 65 nm CMOS Technology Dedicated to Low Power Consumption Wireless Applications

**Raja MAHMOU[1], Khalid FAITAH[2]**

[1] LGECOS, ENSA Marrakech, Cadi Ayyad University, Morocco

[2] LGECOS, ENSA Marrakech, Cadi Ayyad University, Morocco

**Abstract**

The present work consists of designing a Single Balanced Mixer (SBM) with the 65 nm CMOS technology, this for a 1.9 GHz RF channel, dedicated to wireless applications. This paper shows; the polarization chosen for this structure, models of evaluating parameters of the mixer, then simulation of the circuit in 65nm CMOS technology and comparison with previously treated.

## I. Introduction

With the multimedia's advent, the aspect of embedded systems, of RF architecture of transmission / reception channel; require a reduction in size and energy consumption with equal performances. In this stage, advanced CMOS technologies are therefore a new way now increasingly studied for the design of RF functions, the transition frequency of CMOS transistors is inversely proportional to the length of the channel, and has, as such, steady progress of lithography.

In a radio-frequency wireless, especially in a superheterodyne architecture, the frequency mixer is an indispensable module which the impact is critical on the performance of all functions [1].

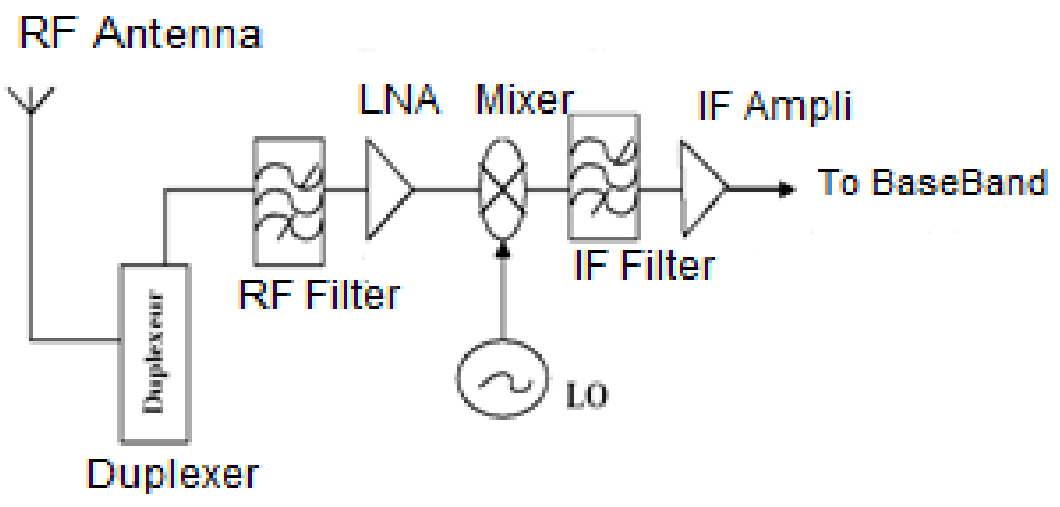

Fig.1 Functional structure of an RF receiver chain

This paper describes a work which develops the design of a single balanced mixer (SBM) with a 65 nm CMOS technology, this for a 1.9 GHz RF channel, dedicated to wireless applications, starting with the polarization of the circuit, calculations of evaluating parameters mixer, then a simulation of the circuit with discussion of results, and finally a potential comparison with the technologies already adopted.

## II. Modeling of evaluating mixer 'Parameters

### 2.1 Architecture of the SBM Mixer

The architecture of the mixer design, shown in Fig.2, is of the type single balanced (SBM).

This structure requires that transistors CMOS M2 and M3 must be identical. The CMOS M1 receives the RF signal and acts as the voltage / current converter. The current trail Is is shared equally among the coupled sources M2 and M3. Thus, the $V_{RF}$ signal varies the drain-source current of M1, and the switching operation of M2 and M3 multiplies this variation by the $V_{LO}$ signal coming from a local oscillator. Finally, the output signal $Vout$ is represented by the voltage between the drains of CMOS M2 and M3 [2].

We have chosen sizing for CMOS 65 nm technology (M1, M2, M3, M4) [3].

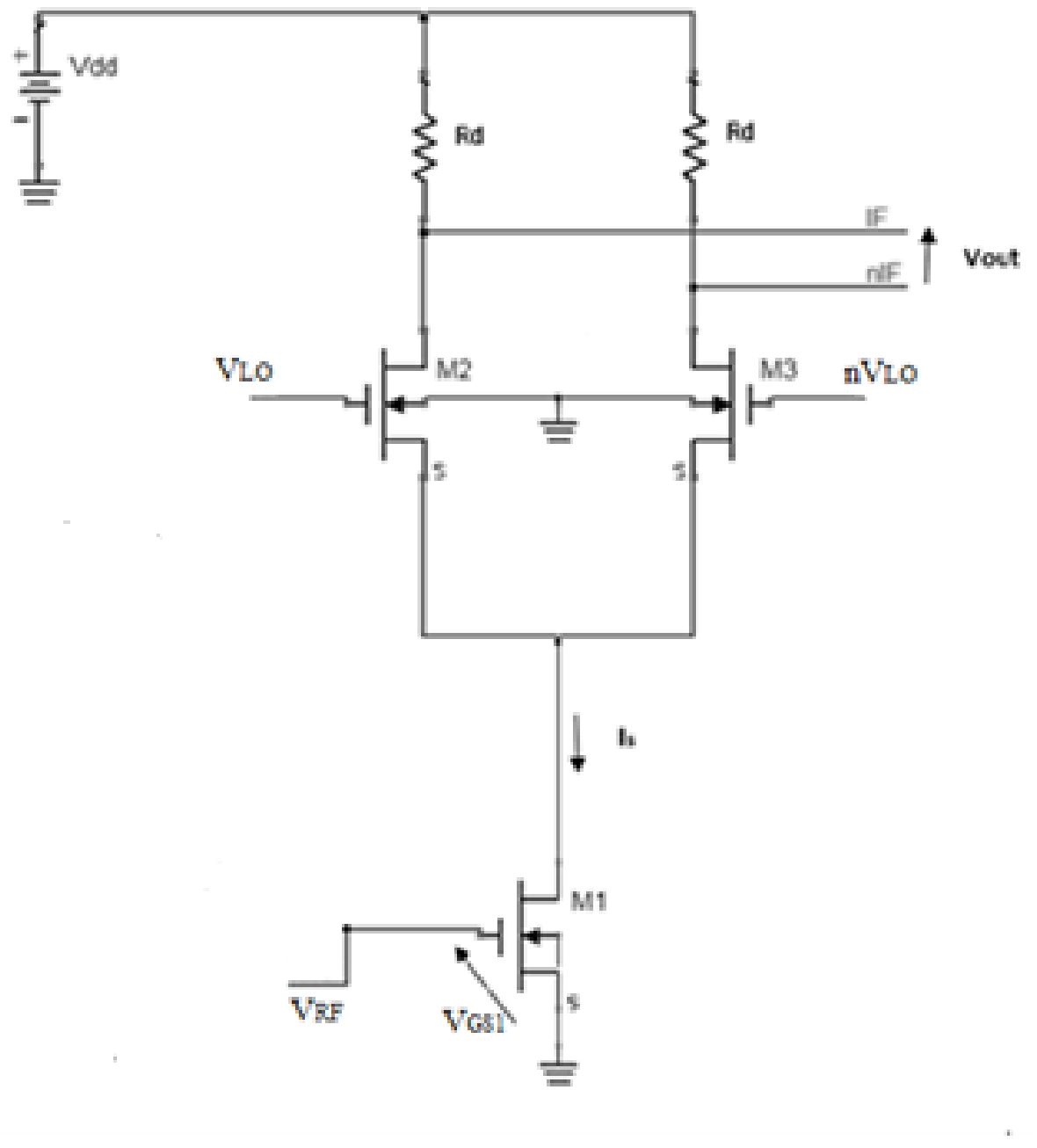

Fig.2 Polarization Diagram of a SBM

$V_{RF}$ and $V_{LO}$ frequencies are respectively 1.9 GHz and 1.8 GHz which provides an intermediate frequency of 100 MHz. The choice of these values gives an IF frequency as agreed to meet most of the wireless networks deployed today, and operating frequency around 1 GHz, such as GSM [4].

## 2.2 Conversion Gain

The chosen architecture is an architecture single balanced with the two CMOS M2 and M3 of the differential pair are on commutation mode (Fig. 2), so the output current is controlled by the state of the VCO signal generated by the local oscillator which allows write:

$$Iout(t) = Is(t).signe[V_{LO}(t)] \quad (1)$$

With $signe[V_{LO}(t)]$ is the Fourier transform of the VCO signal:

$$signe[V_{LO}(t)] = \frac{4}{\pi}\left\{cos(\omega_{LO}t) - \frac{1}{3}cos(3\omega_{LO}t) + \frac{1}{5}cos(5\omega_{LO}t) + ..\right\} \quad (2)$$

According to the circuit and the internal structure of CMOS M1 we have:

$$Is(t) = g_m V_{GS1} + g_m V_{RF} cos(\omega_{RF}t) \quad (3)$$

$g_m$ is the Transductance of the CMOS M1
$V_{GS1}$ is the bias voltage of the M1 CMOS gate, and is the DC component of the $V_{RF}$ signal
Then we obtain:

$$Iout(t) = \{g_m V_{GS1} + g_m V_{RF} cos(\omega_{RF})\}\frac{4}{\pi}\left\{cos(\omega_{LO}t) - \frac{1}{3}cos(3\omega_{LO}t) + \frac{1}{5}cos(5\omega_{LO}t) + \cdots\right\} \quad (4)$$

And as $Vout(t) = R_d\, Iout(t)$ we can write:

$$Vout(t) = \left\{\frac{4 g_m V_{GS1}}{\pi} R_d cos(\omega_{LO}t) + \frac{2}{\pi} R_d g_m V_{RF}\left[cos\left((\omega_{RF} - \omega_{LO})t\right) - cos\left((\omega_{RF} + \omega_{LO})t\right)\right] + \cdots\right\} \quad (5)$$

The conversion gain is [5]:

$$G_{conv} = \frac{V_{out\,\grave{a}t(\omega_{RF}-\omega_{LO})}}{V_{RF}(t)_{\grave{a}t(\omega_{RF})}} = \frac{2}{\pi} R_d g_m \quad (6)$$

On the ADS2009 tool we have chosen a model for CMOS 65nm technology [7]. According to a simplified modeling of the internal structure of the transistor M1, we found $g_m$ which is in the range of 34mA / V; this gives a theoretical conversion gain equal to 13.55dB.

## 2.3 Non-linearity

- **1 dB Compression Point**

Like any electronic device with nonlinears active components, the mixer has an output power curve based on that of the entry and presenting a saturation zone. It is characterized by the 1 dB compression point, defined as the RF input power for which the conversion gain is reduced by 1 dB [6] (Fig.3).

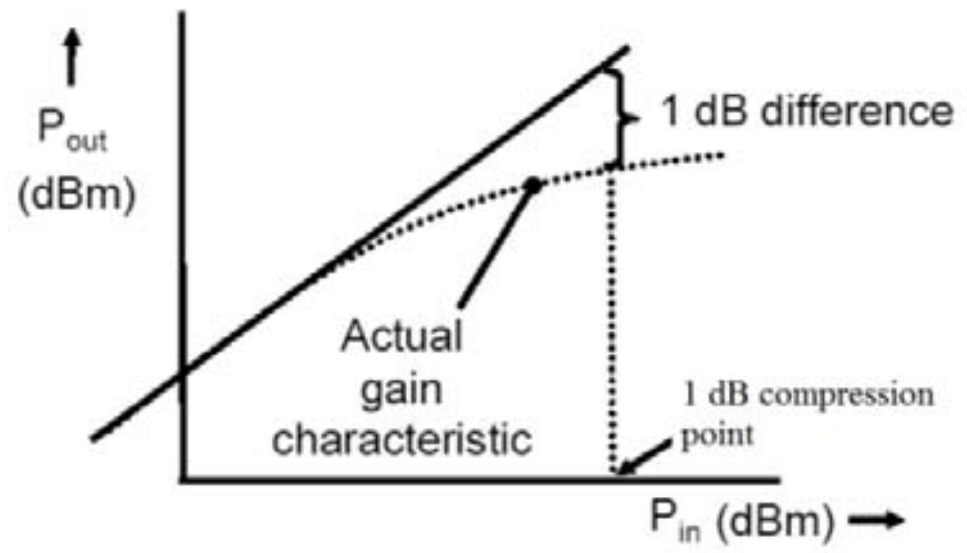

Fig.3 1dB Compression Point

- **Order 3 Interception Point (IIP3)**

Also called the intermodulation level of order 3, characterizes the distortion of the system, in effect: we can notice that around two useful rays f1 and f2, can be superimposed two other very close rays (2.f1 - f2) and (2 . f2 - f1) that can't be easily filtered out (Fig.4) [6].

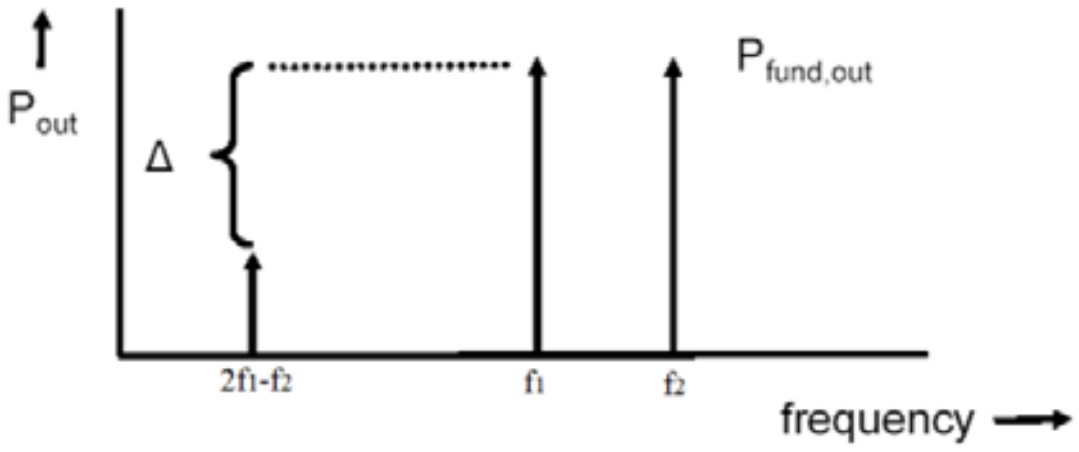

Fig.4 Order 3 interception point

IIP3 is given by the relation:

$$\mathrm{IIP3} = \frac{\Delta}{2} + \mathrm{Power_{RF}} \quad (7)$$

or $\mathrm{Power_{RF}}$ is the input power.

- **Isolation**

Isolation is translated by the power coupled from one port to another. In general for a mixer, whatever its type, isolation is the most critical between RF and LO ports because of their closest frequencies and therefore difficult to filter [5].

$I_{OL_RF}$ will represent the ratio between the local oscillator's power on the channel RF and local oscillator's power injected into the mixer.

$$I_{OL_RF} = \frac{P_{OL_RF}}{P_{OL_OL}} \quad (8)$$

- **Noise Figure**

The noise figure NF of a mixer is defined conventionally as the degradation of signal to noise ratio between the input and the output [5]:

$$F = \frac{\frac{P_{RF}}{N_{RF}}}{\frac{P_{IF}}{N_{IF}}} \quad so \quad F = \frac{N_{IF}}{N_{RF}.G_{conv}} \quad (9)$$

# III. Simulation results

$V_{RF}$ and $V_{LO}$ frequencies are respectively 1.9 GHz and 1.8 GHz which provides to an intermediate frequency IF of 100 MHz.

## 3.1 Transient signals

The Figure 5 show the shape of the IF output signal whose frequency is 100 MHz.

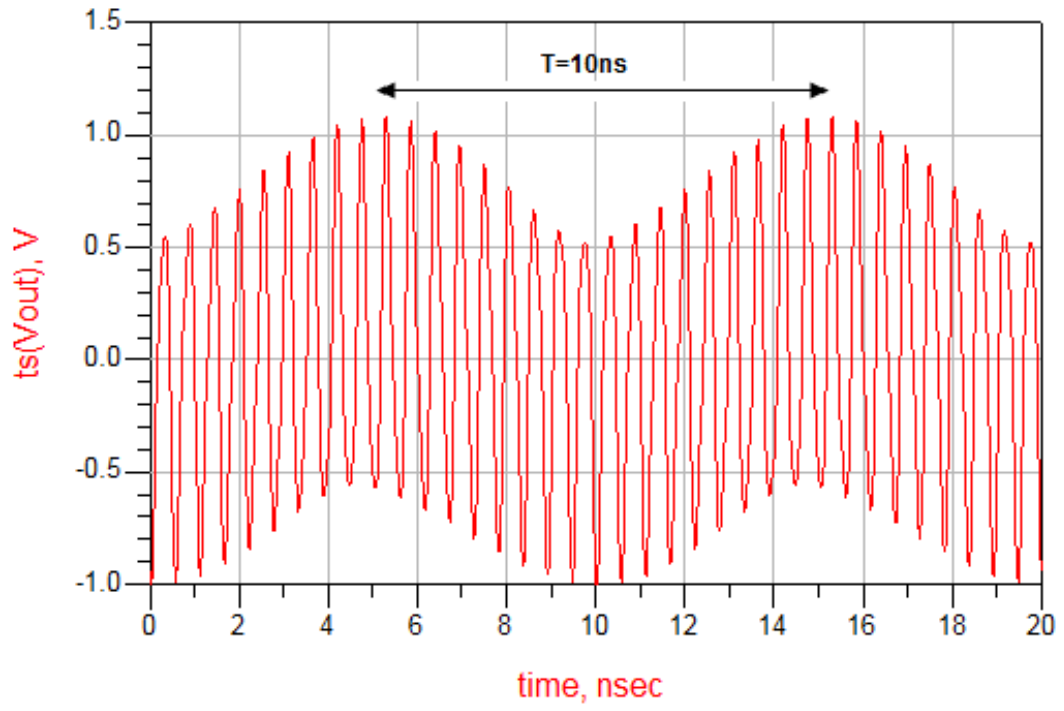

Fig.5 Time response of the output signal *Vout (t)*

Such a chronogram represents in fact the RF carrier and the useful signal IF that we will have to restore it after an adequate filter. The following figure shows *V(t)*, it's the output signal *Vout (t)* after inserting a filter .

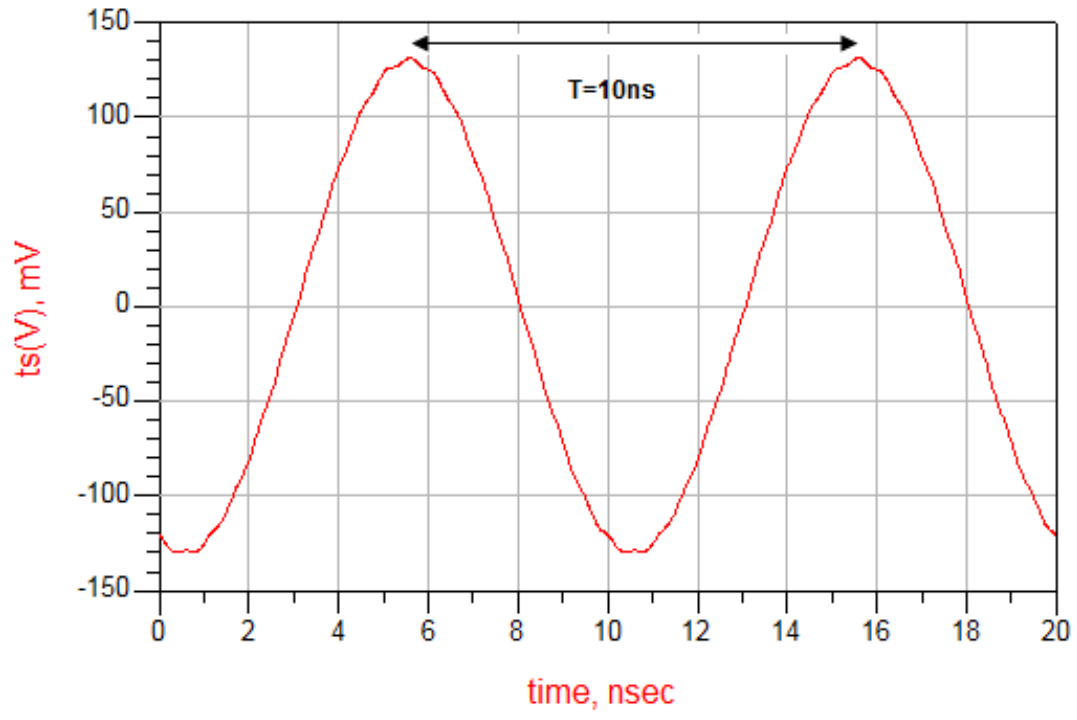

Fig.6: Response time of the output signal *Vout (t)* filtered

## 3.2 Power Consumption

DC simulation, allowed us to measure the power consumption of the mixer circuit which is 2 mW, with Vdd = 1.8V.

## 3.3 Harmonic responses

The Figures 7 and 8 show the harmonics response of order 5 of $V_{RF}$ and $V_{out}$ signals whose basic rays are represented respectively by 1.9 GHz and 100 MHz frequencies.

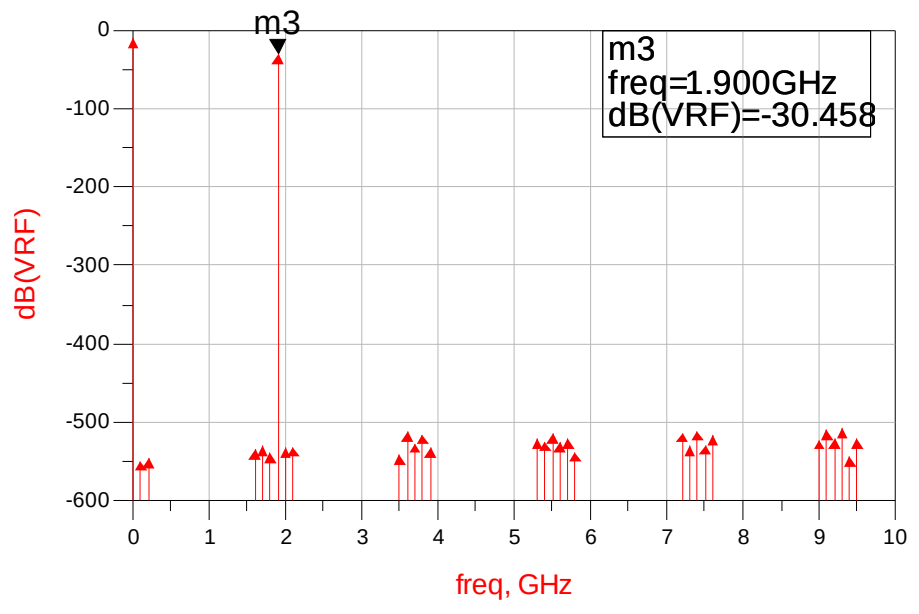

Fig.7 $V_{RF}$ harmonic response

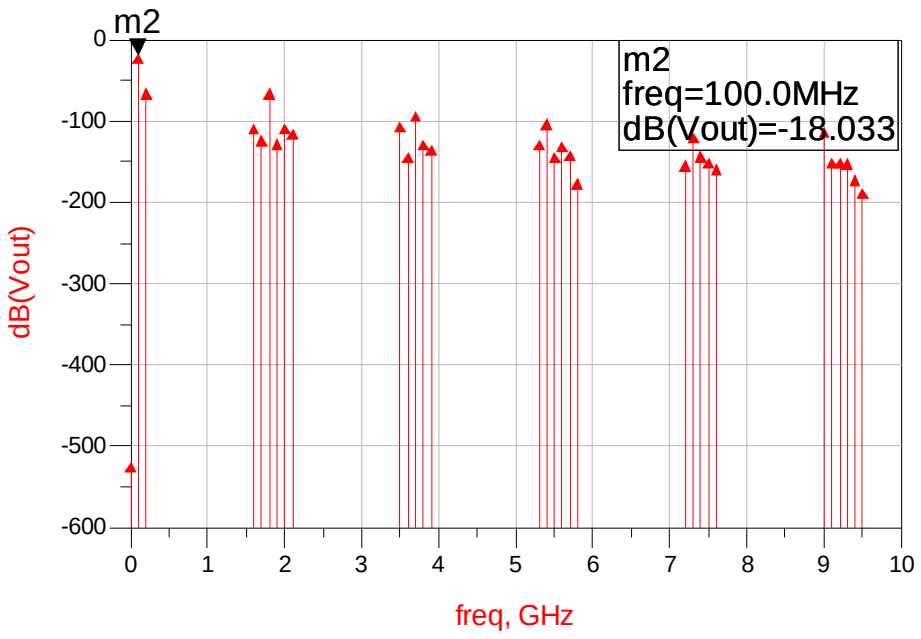

Fig.8 $V_{out}$ harmonic response

Harmonic simulation of mixer circuit results in a conversion gain (equation 6) equal to (-18.033dB)-(-30.458dB) =12.425dB (figures 7 and 8). This reveals the importance of taking into consideration all the internal CMOS parameters which are liable to affect the results.

## 3.4 dBm Gain

We obtain the function shown in Fig.9, which represents a linear zone (horizontal) where the output is directly proportional to the input, and an area decreasing from -11.5dBm resulted by the non-linearity of the mixer circuit.

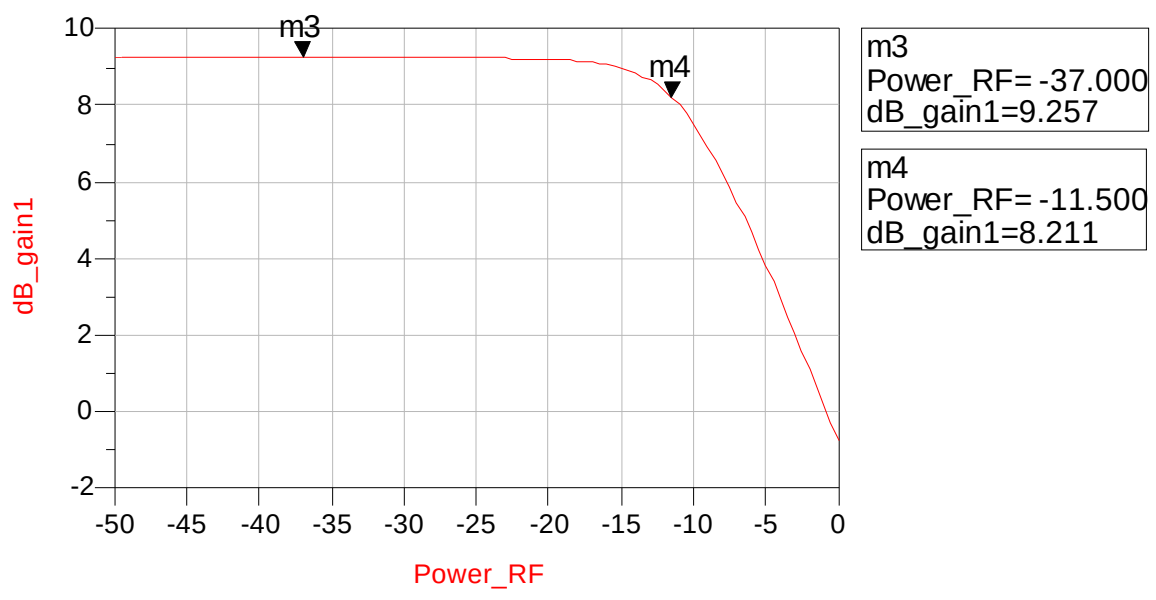

Fig.9: Gain function of the RF Power Input

## 3.5 1 dB Compression Point

As shown in Figure 10, it's the gain value for which the output power (Vout_dBm1) does not follow its right line (Line1) with a difference of 1 dB. It corresponds well to an RF power equal to -11.5 dB.

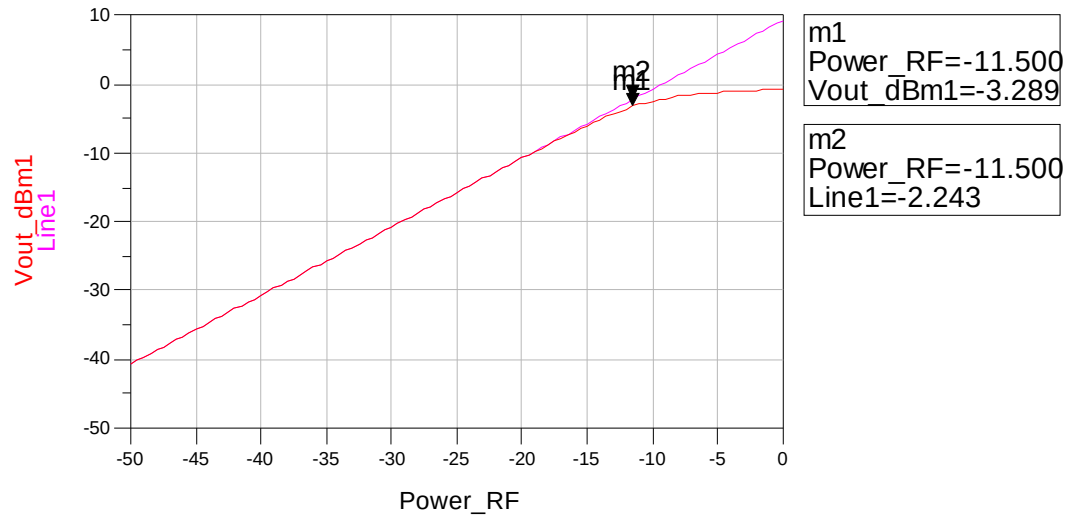

Fig.10 Response $V_{out}$ in dBm function of the Power Input

## 3.6 Interception Point Order 3 (IIP3)

The simulation result shown in Fig.11, and (equation 7) lead to an IIP3 value in the order of 6 dBm.

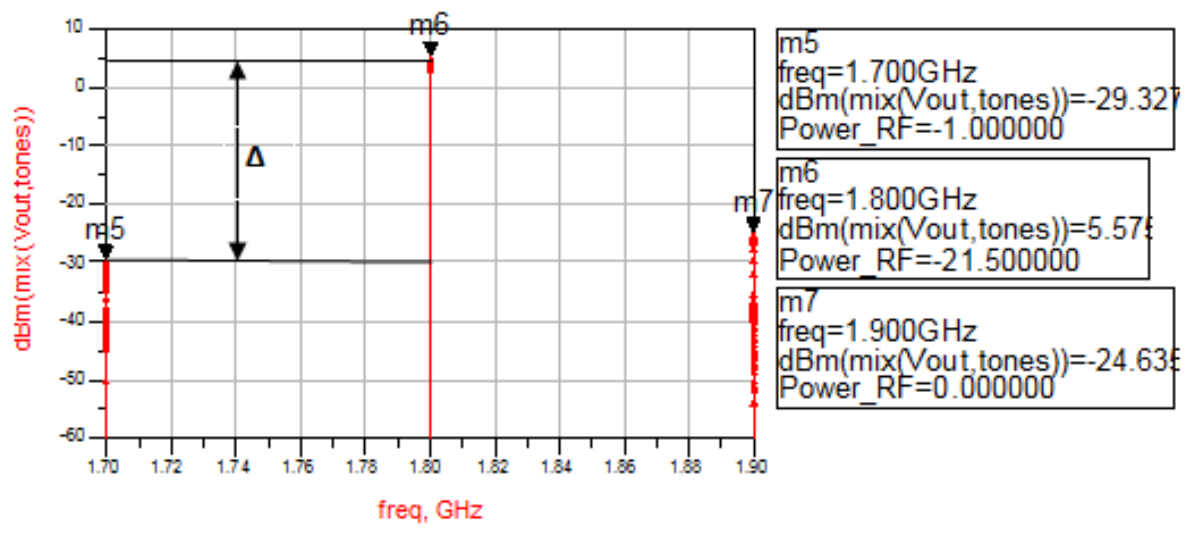

Fig.11 Interception Point IIP3

## 3.7 Measure of Isolation between ports

Isolation report (equation 8) is represented in Fig.12; it is equal to -37.704dB for 1.9 GHz RF frequency

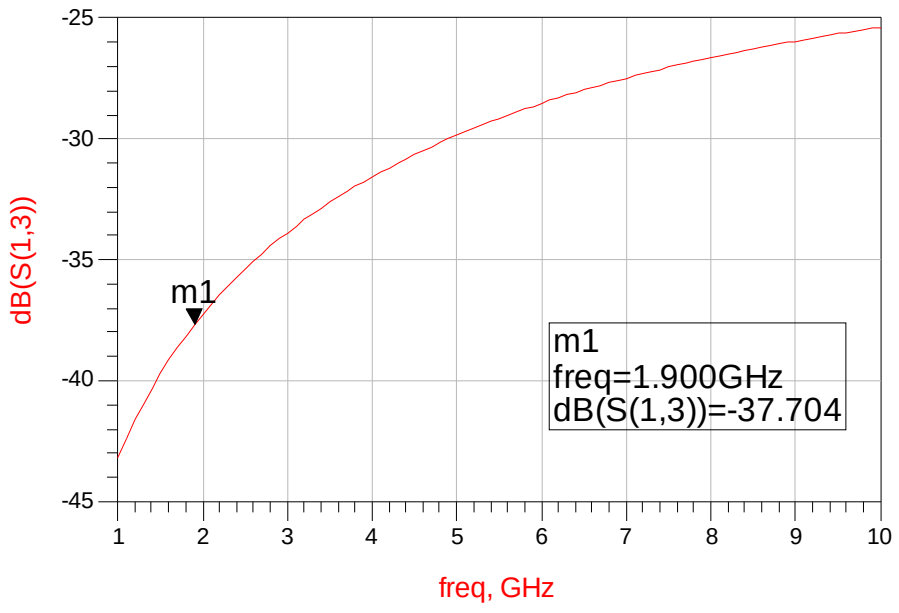

Fig.12 Isolation equal to -37.704 dB for 1.9 GHz RF frequency

### 3.8 Noise Figure

Curve noise in the input and in the output, are shown in the following figures (figures 13 and 14):

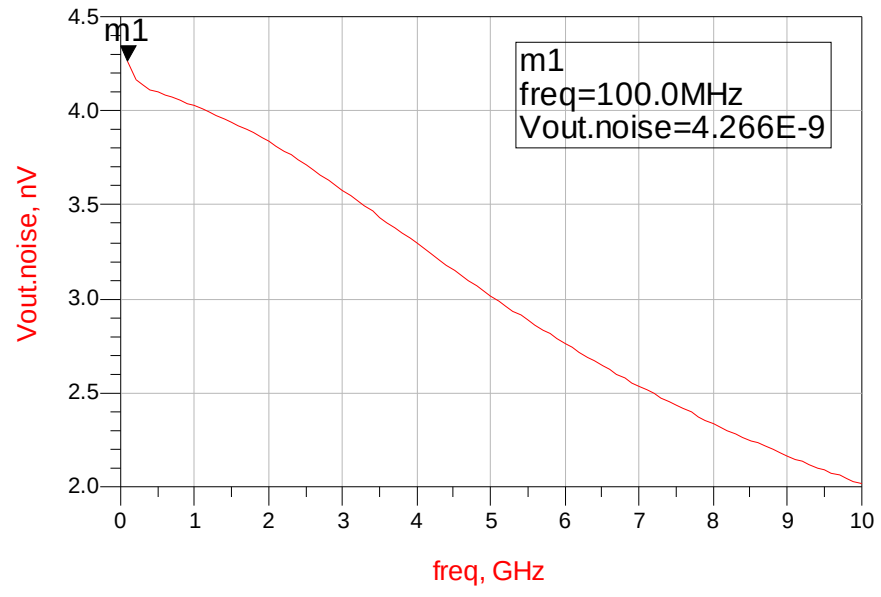

Fig.13 Noise Input

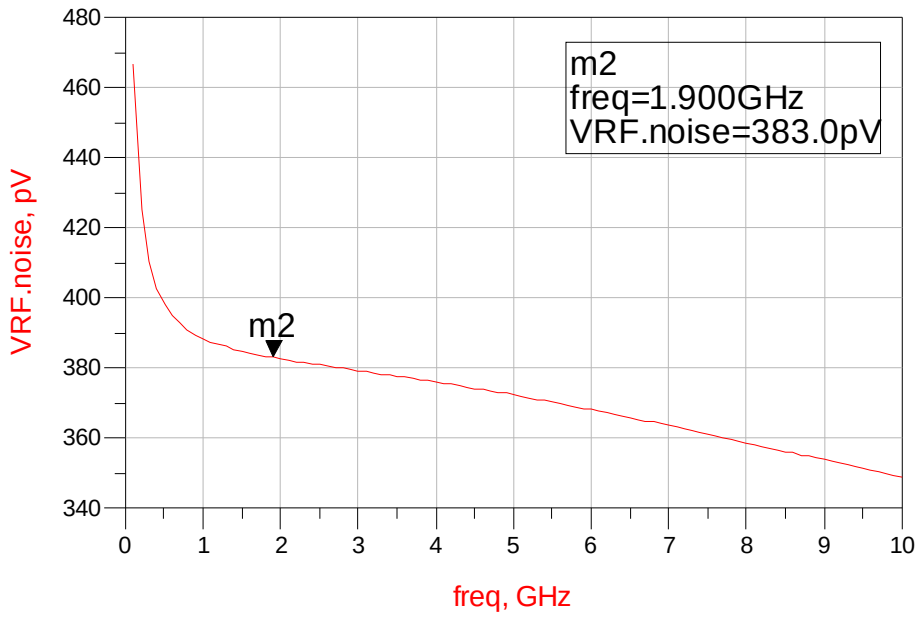

Fig.14 Noise Output

From the curves we obtain: *NIF=4.266nV* and *NRF=0.383nV* and knowing *Gconv = 12.425dB*, by the relation (9) the noise figure is 8.92dB.

## IV. Comparison of performance obtained with recent mixers

According to the simulation results found, choosing the 65nm CMOS technology with the design of other parameters (Rd, Z, .. etc..) of the SBM circuit, allowed us to achieve a very stable gain and linearity over a wide range of input power.

The performances of this mixer (SBM) are compared in the table below with that of some recent mixers:

Table 1: Comparison of Performance Obtained With Recent Mixers

| Réf | Technology (µm) | RF (GHz) | CG (dB) | NF (dB) | P-1 (dBm) | IIP3 (dBm) | Pcons (mW) |
|---|---|---|---|---|---|---|---|
| [8] | 0.35 | 0.9 | 1.1 | - | -15.4 | -3.3 | 7.2 |
| [9] | 0.25 | 2.44 | -2.6 | 13.67 | 5.07 | 12.81 | 13.3 |
| [10] | 0.18 | 2.4 | 3.3 | 14.8 | -8.98 | 5.46 | 5.6 |
| [1] | 0.18 | 1.9 | 7 | 8 | -10 | -5 | 3.8 |
| **Proposed Circuit** | 0.065 | 1.9 | 12.42 | 8.92 | -11.5 | 6 | 2 |

The dynamics of an electronic circuit is defined as the power range for which the functioning is satisfactory. For lower levels, the limitation is set by the noise floor. For high levels, the limiting phenomenon is the compression. Therefore the dynamics of a mixer will be as greatest as its Intercept Point of Order 3 and its 1 dB compression point are important.

We note that the CMOS 65 nm technology SBM design witch we proposed is performing well in terms of Conversion Gain, Power Consumption, levels of IIP3 point and 1 dB compression point, noise figure still acceptable.

## V. Conclusion

The research work presented in this paper is part of the overall objective; to study the feasibility of a Single Balanced Mixer (SBM) in a RF chain, dedicated to wireless applications; by 65nm CMOS technology, also to see from the simulation results, the performance of this choice compared to recent technologies, and finally to proceed to the implementation of this choice.

## References

[1] K. Faitah, A. El Oualkadi, A. Ait Ouahman : "CMOS RF down-conversion mixer design for low-power wireless communications" ACM Ubiquity, Volume 9, Issue 24 June 17 – June 23, 2008.

[2] M. Kramar, S. Spiegel, F. Ellinger, G. Boeck, "A Broadband Folded Gilbert-Cell CMOS Mixer", 14th IEEE International, Conference on Electronics, Circuits and Systems, ICECS 2007, December 11-14, 2007. Marrakech, Morocco.

[3] B.Martineau, "Millimeter-Wave building blocks design methodology in CMOS 65nm process using Agilent tools", ADS users's group meeting, June 16th 2009. STMicroelectronics-Crolles/Minatec.

[4] B. Razavi, "Design considerations for direct conversion receivers", IEEE Transactions on Circuits and Systems-II, Analog and Digital Signal Processing, vol.44, pp: 428-435 (1997).

[5] M.Villegas, "Conception de Circuits RF et micro-ondes", Radio Communications Numeriques/2, 2ème édition, pp : 219 222, Dunod 2002-2007

[6] K.Faitah, "Design of High-Frequency Integrated Transceiver Front-end", ENSA Marrakech, pp 29-30, 39-40, 2008-2009

[7] T.Gneiting, "BSIM4, BSIM3v3 and BSIMSOI RF MOS Modeling", Agilent EEsof EDA Seminar, April 04, 2001 Mountain View, CA, USA.

[8] C.F Au-Yeung and K.K.M.Cheng, "CMOS mixer Linearization by the low frequency signal injection method," IEEE MTT-S International Microwave Symposium Digest, vol 1, pp: 95-98, June (2003)

[9] Kumar Munusamy and Zubaida Yusoff "A Highly Linear CMOS Down Conversion Double Balanced Mixer" ICSE 2006, Kuala Lumpur, Malaysia, pp: 985-990, (2006).

[10] Hung-Che Wei, Ro-Min Weng, Chih-Lung Hsiaoand Kun-Yi Lin, " A 1.5V 2.4GHz CMOS \ Mixer With High Linearity", The 2004 IEEE Asia-Pacific Conference Circuit and Systems, pp: 6-9, Dec, (2004).

**Raja MAHMOU:** Received Professional License degree of Electrical Engineering on July 2007, then Master of Electrical Engineering on July 2009, from the Faculty of Sciences and Technics of Marrakech, at Cadi Ayyad University, Marrakech, Morocco. Now, PhD candidate on Specialty Conception of Analog Systems and RFIC at the Laboratory Of Electrical Engineering and Control Systems (LGECOS) in the National Institute of Applied Sciences (ENSA), Marrakech, Morocco, also actually a Visiting Professor of Analog Electronic, RF CMOS Designs at ENSA Marrakech.

**Dr. Khalid FAITAH:** was born in Rabat, Morocco, in June 1965. He received B.S degree in electronics from Mohamed V University of science, Rabat, Morocco in 1988 and the M.S degree in signal processing in 1997 from Hassan II University of science, Casablanca, Morocco. He received Ph.D degree in electronic at the Ibn Tofal University of science, Knitra, Morocco in 2003. In 2009 he graduated from HDR (Certificate of Accreditation to the search direction). He is now Professor at ENSA (National Institute of Applied Sciences), Department of Electrical Engineering in Cadi Ayyad University, Marrakech, Morocco where he teaches analog electronics, RF CMOS design, sensors and interface circuits and also is Responsible of the Electrical Engineering and Control Systems Laboratory, His research interests include signal integrity and analog RF CMOS design. He is the author/co-author of several publications and communications in recognized journals and international conferences.